\documentstyle[11pt]{article}

\catcode`\@=11
\def\@citex[#1]#2{%
\if@filesw \immediate \write \@auxout {\string \citation {#2}}\fi
\@tempcntb\m@ne \let\@h@ld\relax \def\@citea{}%
\@cite{%
  \@for \@citeb:=#2\do {%
    \@ifundefined {b@\@citeb}%
      {\@h@ld\@citea\@tempcntb\m@ne{\bf ?}%
      \@warning {Citation `\@citeb ' on page \thepage \space undefined}}%
      {\@tempcnta\@tempcntb \advance\@tempcnta\@ne%
      \@tempcntb\number\csname b@\@citeb \endcsname \relax%
      \ifnum\@tempcnta=\@tempcntb 
        \ifx\@h@ld\relax%
          \edef \@h@ld{\@citea\csname b@\@citeb\endcsname}%
        \else%
          \edef\@h@ld{\ifmmode{-}\else--\fi\csname b@\@citeb\endcsname}%
        \fi%
      \else
        \@h@ld\@citea\csname b@\@citeb \endcsname%
        \let\@h@ld\relax%
      \fi}%
    \def\@citea{,\penalty\@highpenalty\,}%
  }\@h@ld
}{#1}}

\def\@citeb#1#2{{[#1]\if@tempswa , #2\fi}}
\def\@citeu#1#2{{$^{#1}$\if@tempswa , #2\fi }}
\def\@citep#1#2{{#1\if@tempswa , #2\fi}}

\def\bcites{         
        \catcode`\@=11
        \let\@cite=\@citeb
        \catcode`\@=12
}

\def\upcites{         
        \catcode`\@=11
        \let\@cite=\@citeu
        \catcode`\@=12
}

\def\plaincites{      
        \catcode`\@=11
        \let\@cite=\@citep
        \catcode`\@=12
}

\newcount\hour
\newcount\minute
\newtoks\amorpm
\hour=\time\divide\hour by 60
\minute=\time{\multiply\hour by 60 \global\advance\minute by-\hour}
\edef\standardtime{{\ifnum\hour<12 \global\amorpm={am}%
        \else\global\amorpm={pm}\advance\hour by-12 \fi
        \ifnum\hour=0 \hour=12 \fi
        \number\hour:\ifnum\minute<10 0\fi\number\minute\the\amorpm}}
\edef\militarytime{\number\hour:\ifnum\minute<10 0\fi\number\minute}

\def\draftlabel#1{{\@bsphack\if@filesw {\let\thepage\relax
   \xdef\@gtempa{\write\@auxout{\string
      \newlabel{#1}{{\@currentlabel}{\thepage}}}}}\@gtempa
   \if@nobreak \ifvmode\nobreak\fi\fi\fi\@esphack}
        \gdef\@eqnlabel{#1}}
\def\@eqnlabel{}
\def\@vacuum{}
\def\marginnote#1{}
\def\draftmarginnote#1{\marginpar{\raggedright\scriptsize\tt#1}}
\def\draft{
        \pagestyle{plain}
        \overfullrule=2pt
        \oddsidemargin -.5truein
        \def\@oddhead{\sl \phantom{\today\quad\militarytime} \hfil
        \smash{\Large\sl DRAFT} \hfil \today\quad\militarytime}
        \let\@evenhead\@oddhead
        \let\label=\draftlabel
        \let\marginnote=\draftmarginnote
        \def\ps@empty{\let\@mkboth\@gobbletwo
        \def\@oddfoot{\hfil \smash{\Large\sl DRAFT} \hfil}
        \let\@evenfoot\@oddhead}
        \def\@eqnnum{(\theequation)\rlap{\kern\marginparsep\tt\@eqnlabel}%
        \global\let\@eqnlabel\@vacuum}  }

\def\blackfonts{
        \font\blackboard=msbm10 scaled\magstep1
        \font\blackboards=msbm8
        \font\blackboardss=msbm6
}

\def\prep{         
        \catcode`\@=11
        \input art10.sty
        \catcode`\@=12
        
        \let\small\null
        \def\blackfonts{
                \font\blackboard=msbm10
                \font\blackboards=msbm7
                \font\blackboardss=msbm5
        }
        \let\sl\it
        \twocolumn
        \sloppy
        \voffset=-2.54truecm
        \hoffset=-2.54truecm
        \flushbottom
        \parindent 1em
        \leftmargini 2em
        \leftmarginv .5em
        \leftmarginvi .5em
        \marginparwidth 48pt
        \marginparsep 10pt
        \setlength{\columnsep}{2truecm}
        \setlength{\textwidth}{25.4truecm}
        \setlength{\textheight}{17truecm}
        \baselineskip=16pt
        \oddsidemargin .18truein
        \evensidemargin .17truein
}

\def\eqalign#1{\null\,\vcenter{\openup\jot\m@th
  \ialign{\strut\hfil$\displaystyle{##}$&$\displaystyle{{}##}$\hfil
      \crcr#1\crcr}}\,}
\def\eqalignno#1{\displ@y \tabskip\centering
  \halign to\displaywidth{\hfil$\@lign\displaystyle{##}$\tabskip\z@skip
    &$\@lign\displaystyle{{}##}$\hfil\tabskip\centering
    &\llap{$\@lign##$}\tabskip\z@skip\crcr
    #1\crcr}}

\def\section{\@startsection {section}{1}{\z@}{3.ex plus 1ex minus
 .2ex}{2.ex plus .2ex}{\large\bf}}
\def\subsection{\@startsection{subsection}{2}{\z@}{2.75ex plus 1ex minus
 .2ex}{1.5ex plus .2ex}{\bf}}        

\def\appendix{{\newpage\section*{Appendix}}\let\appendix\section%
        {\setcounter{section}{0}
        \gdef\thesection{\Alph{section}}}\section}

\def\abstract{\if@twocolumn
\section*{Abstract}
\else 
\begin{center}
{\bf Abstract\vspace{-.5em}\vspace{0pt}}
\end{center}
\quotation
\fi}

\catcode`\@=12

\def\d{\partial}

\def\sqr#1#2{{\vcenter{\vbox{\hrule height.#2pt\hbox{\vrule width.#2pt 
height#1pt \kern#1pt \vrule width.#2pt}\hrule height.#2pt}}}}

\def\=d{\,{\buildrel\rm def\over =}\,}

\def\i3p{\p32\int d^3p}

\def\As{A\hbox to 1pt{\hss /}}
\def\np4{\int d^4p_1\cdots d^4p_{n-1}\, }

\def\Tr{{\rm Tr}\, }

\def\nx4{\int d^4x_1\ldots d^4x_n\, }

\def\kon#1#2{\vbox{\halign{##&&##\cr
\lower4pt\hbox{$\scriptscriptstyle\vert$}\hrulefill &
\hrulefill\lower4pt\hbox{$\scriptscriptstyle\vert$}\cr $#1$&
$#2$\cr}}}

\def\konv#1#2#3{\hbox{\vrule height12pt depth-1pt}
\vbox{\hrule height12pt width#1cm depth-11.6pt}
\hbox{\vrule height6.5pt depth-0.5pt}
\vbox{\hrule height11pt width#2cm depth-10.6pt\kern5pt
      \hrule height6.5pt width#2cm depth-6.1pt}
\hbox{\vrule height12pt depth-1pt}
\vbox{\hrule height6.5pt width#3cm depth-6.1pt}
\hbox{\vrule height6.5pt depth-0.5pt}}
\def\konu#1#2#3{\hbox{\vrule height12pt depth-1pt}
\vbox{\hrule height1pt width#1cm depth-0.6pt}
\hbox{\vrule height12pt depth-6.5pt}
\vbox{\hrule height6pt width#2cm depth-5.6pt\kern5pt
      \hrule height1pt width#2cm depth-0.6pt}
\hbox{\vrule height12pt depth-6.5pt}
\vbox{\hrule height1pt width#3cm depth-0.6pt}
\hbox{\vrule height12pt depth-1pt}}

\def\konw#1#2#3{\hbox{\vrule height12pt depth-1pt}
\vbox{\hrule height12pt width#1cm depth-11.6pt}
\hbox{\vrule height6.5pt depth-0.5pt}
\vbox{\hrule height12pt width#2cm depth-11.6pt \kern5pt
      \hrule height6.5pt width#2cm depth-6.1pt}
\hbox{\vrule height6.5pt depth-0.5pt}
\vbox{\hrule height12pt width#3cm depth-11.6pt}
\hbox{\vrule height12pt depth-1pt}}

\def\i{{\rm int}}

\def\e{{\rm ext}}

\def\a{{\rm av}}

\def\m3{{\mu_1\mu_2\mu_3}}

\def\co{{\rm Com}}

\def\p{{(+)}}

\def\be{\begin{equation}}       \def\eq{\begin{equation}}
\def\ee{\end{equation}}         \def\eqe{\end{equation}}

\def\bea{\begin{eqnarray}}      \def\eqa{\begin{eqnarray}}
\def\ena{\end{eqnarray}}        \def\eea{\end{eqnarray}}
                                \def\eqae{\end{eqnarray}}

\def\ba{\begin{array}}
\def\ea{\end{array}}
\def\unit{1 \hskip-.3em \raise2pt\hbox{$ \scriptstyle |$ } }

\def\a{\alpha}
\def\b{\beta}
 
\def\d{\delta}
\def\e{\epsilon}           
\def\f{\phi}               
\def\g{\gamma}
   
\def\i{\iota}

\def\k{\kappa}                    
\def\l{\lambda}
\def\m{\mu}
\def\n{\nu}
  
\def\p{\pi}                
\def\t{\tau}

\def\D{\Delta}

\def\G{\Gamma}

\def\cd{{\cal D}}

\def\ck{{\cal K}}
\def\cl{{\cal L}}

\def\cn{{\cal N}}
\def\co{{\cal O}}
\def\cp{{\cal P}}

\def\car{{\cal R}}

\def\half{{1 \over 2}}

\def\bop#1{\setbox0=\hbox{$#1M$}\mkern1.5mu
        \vbox{\hrule height0pt depth.04\ht0
        \hbox{\vrule width.04\ht0 height.9\ht0 \kern.9\ht0
        \vrule width.04\ht0}\hrule height.04\ht0}\mkern1.5mu}
\def\Box{{\mathpalette\bop{}}}                        
\def\pa{\partial}                              
\def\ddg{\sp\ddagger} 

\def\>{\rangle} 

\def\<{\langle} 
\def\Dsl{D \hskip-.6em \raise1pt\hbox{$ / $ } }

\def\sl#1{\rlap{\hbox{$\mskip 1 mu /$}}#1}
\def\leftrightarrowfill{$\mathsurround=0pt \mathord\leftarrow \mkern-6mu
       \cleaders\hbox{$\mkern-2mu \mathord- \mkern-2mu$}\hfill
       \mkern-6mu \mathord\rightarrow$}
\def\dvec#1{\vbox{\ialign{##\crcr
       \leftrightarrowfill\crcr\noalign{\kern-1pt\nointerlineskip}
       $\hfil\displaystyle{#1}\hfil$\crcr}}}          
\def\hook#1{{\vrule height#1pt width0.4pt depth0pt}}
\def\leftrighthookfill#1{$\mathsurround=0pt \mathord\hook#1
       \hrulefill\mathord\hook#1$}
\def\underhook#1{\vtop{\ialign{##\crcr                 
       $\hfil\displaystyle{#1}\hfil$\crcr
       \noalign{\kern-1pt\nointerlineskip\vskip2pt}
       \leftrighthookfill5\crcr}}}
\def\smallunderhook#1{\vtop{\ialign{##\crcr      
       $\hfil\scriptstyle{#1}\hfil$\crcr
       \noalign{\kern-1pt\nointerlineskip\vskip2pt}
       \leftrighthookfill3\crcr}}}

\def\sfrac#1#2{{\vphantom1\smash{\lower.5ex\hbox{\small$#1$}}\over
       \vphantom1\smash{\raise.4ex\hbox{\small$#2$}}}} 
\def\bfrac#1#2{{\vphantom1\smash{\lower.5ex\hbox{$#1$}}\over
       \vphantom1\smash{\raise.3ex\hbox{$#2$}}}}      
\def\afrac#1#2{{\vphantom1\smash{\lower.5ex\hbox{$#1$}}\over#2}}  
\def\on#1#2{{\buildrel{\mkern2.5mu#1\mkern-2.5mu}\over{#2}}}
\def\ddt#1{\on{\hbox{\LARGE .\kern-2pt.}}#1}             
\def\tdt#1{\on{\hbox{\LARGE .\kern-2pt.\kern-2pt.}}#1}   

\def\boxes#1{
       \newcount\num
       \num=1
       \newdimen\downsy
       \downsy=-1.5ex
       \mskip-2.8mu
       \bo
       \loop
       \ifnum\num<#1
       \llap{\raise\num\downsy\hbox{$\bo$}}
       \advance\num by1
       \repeat}
\def\boxup#1#2{\newcount\numup
       \numup=#1
       \advance\numup by-1
       \newdimen\upsy
       \upsy=.75ex
       \mskip2.8mu
       \raise\numup\upsy\hbox{$#2$}}

\newskip\humongous \humongous=0pt plus 1000pt minus 1000pt
\def\caja{\mathsurround=0pt}
\def\eqalign#1{\,\vcenter{\openup2\jot \caja
       \ialign{\strut \hfil$\displaystyle{##}$&$
       \displaystyle{{}##}$\hfil\crcr#1\crcr}}\,}
\newif\ifdtup

\def\to{\rightarrow}

\def\1ov4{{1\over 4}}

\def\Tr{{\rm Tr}}

\def\pa{\partial}

\def\dda{\dot{\alpha}} 
\def\ddb{\dot{\beta}}

\def\ddd{\dot{\delta}}

\def\ddg{\dot{\gamma}}

\def\ddt{\dot{\t}}

\def\pa{\partial}

\def\nonu{\nonumber \\{}}
\def\half{{1 \over 2}}

\def\pa{\partial}
\renewcommand{\a}{\alpha}
\renewcommand{\b}{\beta}

\renewcommand{\d}{\delta}
\newcommand{\rmd}{{\rm d}}

\newcommand{\beq}{\begin{equation}}
\newcommand{\eeq}{\end{equation}}
\def\ba{\begin{eqnarray}}
\def\ea{\end{eqnarray}}

\begin{document}



\null\vskip-24pt
\hfill SPIN-1999/11
\vskip-10pt
\hfill THES-TP 99/04
\vskip-10pt
\hfill {\tt hep-th/9906030}
\vskip0.3truecm
\begin{center}
\vskip 2truecm
{\Large\bf
A non-renormalization theorem for conformal anomalies
}\\ 
\vskip 1.5truecm
{\large\bf
Anastasios Petkou${}^\dagger$\footnote{
email:{\tt apetkou@deepblue.physics.auth.gr}} 
and Kostas Skenderis${}^\star$\footnote{
email:{\tt K.Skenderis@phys.uu.nl}}
}\\
\vskip 1truecm
${}^\dagger$ {\it Department of Theoretical Physics,
Aristotle University of Thessaloniki, \\
Thessaloniki 54006, Greece}\\
\vskip 0.5truecm
${}^\star$ {\it Spinoza Institute, University of Utrecht,\\
Leuvenlaan 4, 3584 CE Utrecht, The Netherlands}
\vskip 1truemm
\end{center}
\vskip 1truecm
\noindent{\bf Abstract:}

We provide a non-renormalization theorem for the 
coefficients of the conformal anomaly associated 
with operators with vanishing anomalous dimensions.
Such operators include  conserved currents and 
chiral operators in superconformal field 
theories. We illustrate the theorem by computing the conformal anomaly 
of 2-point functions both by a computation in the conformal 
field theory and via the adS/CFT correspondence.
Our results imply that 2- and 3-point functions
of chiral primary operators in  $\cn=4$ $SU(N)$ SYM 
will not renormalize provided that 
a ``generalized Adler-Bardeen theorem'' holds. 
We further show that recent arguments connecting 
the non-renormalizability of the above mentioned
correlation functions 
to a bonus $U(1)_Y$ symmetry are incomplete due 
to possible $U(1)_Y$ violating contact terms. 
The tree level
contribution to the contact terms may be set to 
zero by considering appropriately normalized 
operators. Non-renormalizability of the above
mentioned correlation functions, however, 
will follow only if these contact terms saturate by 
free fields.

\vfill
\vskip4pt

\eject
\newpage

\section{Introduction}

Supersymmetry has been an instrumental tool in probing the
strong coupling regime of string and quantum field theories. 
In many cases, information which is regarded as dynamical
in non-supersymmetric theories follows by symmetry 
considerations in supersymmetric ones. Certain sectors
of the theory are so severely constraint that no quantum
corrections are allowed at all. A famous example
is the non-renormalization theorem of \cite{GRS}
which states that the superpotential does 
not receive any perturbative corrections. 
Another symmetry which, when present, severely constrains the structure 
of field theories is conformal symmetry. 
Conformal invariance has been quite successfully exploited 
in two dimensions where the conformal group is 
infinite dimensional. In more than two dimensions the conformal group is 
finite dimensional, nevertheless it is powerful 
enough to determine 2- and 3-point functions up to constants
\cite{TMP,OP,erdm3}. The aim of the 
present paper is to discuss non-renormalization theorems
that follow from the combination of supersymmetry and
conformal invariance. In particular, we will study  
the trace anomaly in the presence of external sources for
composite operators. We will show that the trace of the
energy-momentum tensor acquires contributions proportional
to the sources and that certain ratios
of the coefficients of these terms to the overall normalizations
of 2- and 3-point functions do not renormalize. 
Recent studies of the constraints
imposed by superconformal invariance can be found in
\cite{HW1,HW5,park1,erdm1,Osb,erdm2,park3,WP}.

One of the motivations for this investigation comes 
from the recently proposed duality \cite{Malda,Gubs,Wit}
between string or M-theory compactifications on $adS_{d+1}$ 
spacetimes and $d$-dimensional superconformal field theories 
(adS/CFT duality).
In particular, IIB string theory on $adS_5 \times S^5$ 
with $N$ units of $F_5$ flux is 
believed to be dual to $\cn=4$, $d=4$ $SU(N)$ SYM theory. 
The 't Hooft coupling constant, $\l=g_{YM}^2 N$, is related 
to the $adS$ curvature and $N$ 
is related to the string coupling constant, such that the strong 
coupling regime of the 't Hooft large $N$ limit of 
$\cn=4$, $SU(N)$ SYM theory (i.e. $N \to \infty$ with $\l$ fixed but large),
has a weakly coupled description in terms of 
anti-de Sitter supergravity. An essential test for any  
strong-weak coupling duality involves quantities
that do not renormalize as one goes from weak to 
strong coupling. Computing such 
quantities in both weak and strong coupling regimes  and finding agreement
only provides a non-trivial consistency check; 
a possible disagreement 
between the two computations would definitely demonstrate the
falsification of the duality conjecture. 
One of the aims of this investigation is to provide 
such non-renormalization theorems, and subsequently 
test the adS/CFT duality. 

Turning things around, one can also use an unexpected 
agreement between quantities computed in the two 
dual theories as an indication of a possible 
non-renormalization theorem. For instance,
using the adS/CFT correspondence
the gravitational trace anomaly was computed in \cite{HS}
and exact agreement was found with the weakly 
coupled (free) field theory calculation,
indicating that there is a non-renormalization theorem
at work. Indeed, such a renormalization 
theorem has been discussed in \cite{Freed,GK,HSW}. 
In \cite{Gubs,Wit}
a prescription was given for calculating 
boundary Green's functions from anti-de Sitter 
supergravity. With this prescription all two-
and three-point functions of chiral 
primary operators were calculated in the large $N$ limit
in \cite{LMRS}. Surprisingly, the result was found to coincide with the
weak coupling (free field 
theory) computation. This led the authors of \cite{LMRS}
to conjecture that the 3-point functions 
of all chiral primary operators for large $N$ are independent
of $\l$ and to speculate that they might be independent 
of $g_{YM}$ even for finite $N$. Subsequent field theory 
calculations \cite{HFS,GKP} supported the conjecture 
in its strongest form. In \cite{Intr1,Intr2,EHW}
the non-renormalization of the 2- and 3- point function
was claimed to follow from a bonus $U(1)_Y$ symmetry 
that $\cn=4$ SYM should have in an appropriate limit.
However, as we will discuss, contact term contributions to correlation 
functions render the argument of \cite{Intr1,Intr2} incomplete.
Only if the contact terms saturate by free fields non-renormalization
would follow.
The results of the present  
paper imply that (properly normalized) 2- and 3-point functions 
of chiral primary operators
will be independent of $g_{YM}^2$ for any $N$, if (i)
the standard relation between trace anomalies and 
R-anomalies due to supersymmetry holds in the presence
of sources for composite operators and (ii)
the divergence of the R-current in the presence of external sources 
saturates by free fields. 

This paper is organized as follows. In the next section 
we recall the connection between conformal anomalies 
and short distance singularities. Section 3 contains 
the non-renormalization theorem. In section 4, we use 
the theorem to test the adS/CFT correspondence by computing 
the conformal anomaly associated with 2-point functions.
Section 5 
contains some further comments about the application 
of our results to $\cn=4$ SYM and an analysis of the 
contact terms enter in the discussion of the $U(1)_Y$ 
bonus symmetry. We conclude in section 6.

\section{Conformal Anomalies and Short Distance Singularities}

Consider a CFT in a $d$-dimensional curved background with metric
$g_{\m\n}(x)$ and let its generating functional for connected renormalized
$n$-point functions of a scalar composite operator ${\cal{O}}(x)$ 
be denoted by $W_{R}\equiv W_{R}[g,J]$, where $J(x)$ is an external  source 
for ${\cal{O}}(x)$.  The 
discussion generalizes straightforwardly to the case of many,
possibly non-scalar, composite operators. 
Correlation functions are obtained by 
functional differentiation in a standard fashion, 
\be
\langle{\cal{O}}(x)\rangle =
\frac{1}{\sqrt{g}}\frac{\delta W_{R}}{\delta J(x)}, \qquad
\langle T_{\m\n}(x)\rangle
=\frac{2}{\sqrt{g}}\frac{\delta W_{R}}{\delta g^{\m\n} (x)}\,.\label{sour}
\ee
Requiring that $W_{R}$ does not change under constant Weyl rescalings
of $g_{\m\n}(x)$ and $J(x)$ leads to the following renormalization
group equation (we are at the 
fixed point where 
the beta functions vanish) \cite{Osborn},
\be
\left( \m\frac{\partial}{\partial\m} +\int \rmd^{d}x\left[
-2g^{\m\n}(x)\frac{\delta}{\delta
g^{\m\n}(x)}
+(d-\Delta)J(x)\frac{\delta}{\delta J(x)}\right]\right) W_{R}=0\,,
\label{rge}
\ee
where $\m$ is an arbitrary mass-scale and $\D$ is the conformal
dimension of ${\cal{O}}(x)$.  

Note that (\ref{rge}) is general enough to include the possible existence of
conformal anomalies in the theory. These, as is well known,
are intimately connected with  UV 
renormalization. Their coefficients are quantitatively related to the
subtraction procedure of short-distance singularities in $n$-point
functions. Let us briefly recall this fact.
The  generating functional of connected graphs may be written as
\bea
W_{R}=\sum_{k=1}^{\infty}\frac{1}{k!} \int\rmd^{d}x_{1}\sqrt{g}
\,..\,\rmd^{d}x_{k}\sqrt{g} \,J(x_{1}).. J(x_{k})\,\langle{\cal{O}}(x_{1})
..{\cal{O}}(x_{k})\rangle_{R}\,.\label{eq6}
\eea
The usual renormalization procedure requires that both $J(x)$ and
$\langle{\cal{O}}(x_{1}) 
...\rangle_{R}$ are $\m$-dependent \cite{Zinn-Justin}. Under the Weyl
rescalings considered in  
(\ref{rge}), the scalar sources in (\ref{eq6}) transform as
$\m(\partial/\partial\m)J(x)=-(d-\Delta)J(x)$. Then, we can 
calculate the partial derivative of $W_{R}$ as 
\bea
\mu\frac{\partial}{\partial\mu} W_{R} &= &
 -(d-\Delta)\int\rmd^{d}x\sqrt{g}\,J(x)\,\frac{\delta W_{R}}{\delta
J(x)} \nonumber \\
& & +\sum_{k=1}^{\infty}\frac{1}{k!} \int\rmd^{d}x_{1}\sqrt{g} 
\,..\,\rmd^{d}x_{k}\sqrt{g}
\,J(x_{1})..
J(x_{k})\,\m\frac{\partial}{\partial\mu}\langle{\cal{O}}(x_{1}) 
..{\cal{O}}(x_{k})\rangle_{R}\,. \label{mder}
\eea
By virtue
of (\ref{mder}), we then obtain from (\ref{rge})  
\bea
\sum_{k=1}^{\infty}\frac{1}{k!} \int\rmd^{d}x_{1}\sqrt{g} 
\,..\,\rmd^{d}x_{k}\sqrt{g}
J(x_{1})..
J(x_{k})\m\frac{\partial}{\partial\mu}\langle{\cal{O}}(x_{1}) 
..{\cal{O}}(x_{k})\rangle_{R} =\int\rmd^{d}x\sqrt{g}g^{\m\n}(x)\langle
T_{\m\n}(x)\rangle,\label{eq8}
\eea
which shows that possibly existing local  delta-function terms in
the renormalized $n$-point functions of 
${\cal{O}}(x)$ are related to conformal anomalies. Since the dimension of
$T_{\m\n}(x)$ is $d$, (\ref{eq8}) implies that conformal $n$-point
functions of  
scalar operators ${\cal{O}}_{i}(x)$ with dimensions $\Delta_{i}$,
$i=1,2,..,n$ may in general contribute to the conformal anomaly if
\be
\sum_{i=1}^{n}\Delta_{i}=(n-1)d+2k\,,\,\,\,\,\,\,\,\, k=0,1,2,..\,. \label{eq9}
\ee
For example, the two-point function of a scalar operator with dimension
$\Delta$ contributes to the conformal anomaly when
\be
\Delta=\frac{d}{2}+k\,,\,\,\,\,\,\,\,\, k=0,1,2,..\,.\label{eq10}
\ee

It is well known that conformal invariance fixes the 
form of scalar 2- and 3-point functions up to 
an overall constant \cite{TMP,OP}, 
\bea
 G^{IJ}_{(2)}(x_1,x_2;\D) &=& \< \co^I(x_1) \co^J(x_2)\>=
\frac{C_2^{IJ}(g,\D)}{(x_{12}^2)^{\Delta}}\,, \label{2pt} \\
G_{(3)}^{IJK}(x_{i};\Delta_{i}) &=&  \< \co^I(x_1) \co^J(x_2)
\co^K(x_3)\> \,\nonumber \label{2-pt}\\
&=&
\frac{C^{IJK}_3(g,\D_i)}{[x_{12}^{2}]^{\frac{1}{2}(\Delta_{1}+\Delta_{2}-
\Delta_{3})}[x_{13}^{2}]^{\frac{1}{2}
( \Delta_{1}+\Delta_{3}-\Delta_{2})}[x_{23}^{2}]^{\frac{1}{2}
( \Delta_{2}+\Delta_{3}-\Delta_{1})}}\,, \label{3pt} 
\eea 
where $x^2_{ij}=|x_i-x_j|^2$, $g$ denotes the coupling constant(s)
of the theory and
the operators $\co^{(I,J,K)}$ have dimensions $\D_{(I,J,K)}$.  $C^{IJ}_2$ and 
$C^{IJK}_3$ are constants that may depend on the coupling 
constant(s) and scaling dimensions of the operators involved.
It follows from (\ref{2pt}) and (\ref{3pt}) 
that they are symmetric in their indices. 
The operators $\co^I$ and $\co^J$ in the 2-point function necessarily  
have the same dimension $\D_I=\D_J=\D$, thus one may choose a basis
such that $C_2^{IJ}(g,\D)=\d^{IJ} C_2(g,\D)$. 

One might wonder how is it possible for the conformal field
theory correlation functions to depend on a scale $\mu$ when their 
form has been determined by conformal invariance. The
answer is that the distributions appearing in the 
right hand side of (\ref{2pt}) and (\ref{3pt}) are conformal 
away from coincidence points but, depending on 
the scaling dimension of the operators involved,
they may acquire singularities when some of the points come close together. 
Regularizing such  singularities entails introducing a
mass-scale $\mu$, which in turn allows for the possibility
of conformal anomaly.
If the $\mu$ derivative of the regularized
2- and 3-point (or, more generally, of $n$-point) function
yields a local term, then (as it follows from (\ref{eq8}))
the trace  of the energy momentum tensor
acquires a contribution of the form,
\be
\< T^\m{}_\m \> = \frac{1}{2}\cp_{(2)}^{IJ} (\pa^2)^{k_{IJ}} J^{I} J^J 
+ \frac{1}{3!}\cp_{(3)}^{IJK} (\pa^2)^{k_{IJK}} J^I J^J J^K + \cdots
\,,\label{emtr} 
\ee
where the integers $k_{IJ}$ and $k_{IJK}$ can be read off 
from (\ref{eq9}). One can check, using (\ref{eq9}), that 
both sides of this equation scale the same way, as it should.

\section{The Theorem}

We are now ready to prove the following non-renormalization 
theorem: 

{\bf Theorem:} 
{\it The ratio of the conformal anomaly associated with
2- and 3-point functions
to the overall constant appearing in the 2- respectively 3-point functions
does not renormalize, provided the anomalous dimensions of the operators
involved are zero.}

{\bf Proof:} Since the form of the 2- and 3-point functions 
of operators of given dimensions is uniquely determined by   
conformal invariance, 
2- and 3-point functions of operators with vanishing
anomalous dimensions will be given by (\ref{2pt})
for any value of the coupling constant.
The overall constants $C_2$ and $C_3$ may receive quantum corrections. 
However, because the anomalous dimensions of the operators
involved and the beta function vanish, $C_2$ and $C_3$ 
are renormalization group invariants, i.e.
\be 
\m {\pa \over \pa \mu} C_2 =  \m {\pa \over \pa \mu} C_3 = 0\,.
\ee
Therefore, from (\ref{eq8}) and (\ref{emtr})
\be
{\cp_{(i)} \over C_i} =  \m {\pa \over \pa \mu} {G_{(i)} \over C_i}\,,
\ee
which proves our assertion, since the $\mu$ dependence of 
$G^{(i)}/C_i$ follows from the mathematical properties
of the distribution, the latter being fixed at all scales from
conformal invariance alone. \hfill $\Box$

So far our considerations apply to any conformal field theory. 
Although we have only considered scalar operators,
one can (appropriately) 
generalize the discussion to non-scalar operators as well.
A class of operators with zero anomalous dimensions are 
conserved currents. Another class of such operators are chiral 
operators in superconformal theories \cite{ColWe}. 
When supersymmetry is present, one can often connect 
correlation functions of non-scalar operators 
to ones of only scalar operators. This is, for instance, the case
when $d=4$, $\cn=4$ supersymmetry is present (as we review
in section 5). In these cases, it is supersymmetry that extends the results
of the theorem to non-scalar operators.

In the case of superconformal theories, supersymmetry, 
apart from protecting the dimensions of chiral 
operators, also connects the trace anomaly to the 
divergence of the R-current. Whether this connection 
holds in the presence of composite operators is
an open question. Assuming such a connection,
the non-renormalizability of 2- and 3-point 
functions would follow if the R-current satisfies
a sort of ``generalized Adler-Bardeen theorem''.
By this we mean that the divergence of the 
R-current in the presence of composite operators
will saturate by free fields. We are not aware 
of any discussion, let alone a proof, of such theorem
in the literature. Let us further note that even the 
validity of the standard Adler-Bardeen theorem is 
not automatically guaranteed in the case of superconformal
theories. The issue of the compatibility of such a theorem 
with supersymmetry has been the source of much controversy in the '80s
(see for instance \cite{GMZ,SV}; for a recent discussion 
we refer to  \cite{AHM}). We note, however, that 
the $\cn>1$ cases are less subtle.
The standard proofs of the Adler-Bardeen theorem
(see, for instance, \cite{collins}) 
consist of showing that the coefficient of the anomaly 
is a renormalization group invariant. However, 
when the beta functions are zero, as in the case 
of stationary renormalization group flows, this fact by itself does not 
imply that such a coefficient is independent
of the coupling constant. For the case of $\cn=4$ SYM 
a different argument was given in \cite{HSW} for the 
one loop nature of the chiral anomaly. In our case,
we would like to have such a theorem for the divergence 
of the axial current in the presence of sources for 
composite operators. 

\section{AdS/CFT duality}

In certain cases, a given conformal field theory has a conjectured
adS dual. In these cases, one can use the non-renormalization 
theorem provided in the previous section in order to 
test the adS/CFT duality. As an illustration we 
compute the conformal anomaly associated with 
2-point function using both CFT and adS methods. 
The (considerably more involved) computation 
of the conformal anomaly associated with 3-point functions 
will be presented elsewhere \cite{PS}. 

\subsection{CFT computation}

In this section we compute the conformal anomaly associated
with 2-point functions using differential renormalization
techniques \cite{diffreg}.
2-point functions involve distributions of the form $(x^2)^{-\l}$.
For $x \to 0$ they behave as 
\be \label{sing}
{1 \over (x^2)^\l} \sim {1 \over d + 2k - 2\l} {1 \over 2^{2k}k!}
{\G(\half d) \over \G(\half d +k)} S_{d-1} (\pa^2)^k \d^{(d)}(x)\,,
\ee
where $S_{d-1}=2 \p^{d/2}/\G(d/2)$ is the volume of the unit $(d-1)$-sphere.  
From (\ref{sing}) we see that $(x^2)^{-\l}$
has poles at $\l=\half d +k, k=0,1,...$. To get a well-defined
distribution we need to subtract the pole.
We do this by first considering
arbitrary $\l$, subtracting the pole, and then letting $\l \to d/2 +k$.
It suffices to work out the case $\D=d/2$, since the other 
ones can be obtained by differentiation as
\be \label{D-d2}
\frac{1}{(x_{23}^2)^{\frac{1}{2}d+k}} = 
c_{k} (\partial^2)^{k}\frac{1}{(x^2_{23})^{\frac{1}{2}d}},   \qquad
c_{k}= \frac{\Gamma(\frac{1}{2}d)}{4^{k}\Gamma(k+1)
\Gamma(\frac{1}{2}d+k)}
\ee
Following \cite{OP}, we renormalize $(x^2)^{-d/2}$ 
as follows,
\be \label{subtract}
\car {1 \over (x^2)^{\frac{1}{2}d}} \equiv \lim_{\l \to d/2}
\left({1 \over (x^2)^\l} 
- {\m^{2 \l-d}  \over d-2 \l} S_{d-1} \d^{(d)}(x) \right)
\ee
A short calculation yields
\be \label{d/2ren}
\car {1 \over (x^2)^{\frac{1}{2}d}} = 
-{1 \over 2(d-2)} \pa^2 {1 \over (x^2)^{{\frac{1}{2}d}-1}}\left(\log \m^2 x^2 
+ {2 \over d-2}\right)\,.
\ee
The coefficient of the last term, which is proportional 
to $\d^{(d)}(x)$, is scheme dependent, but its precise value
does not enter in the computation of the conformal anomaly.

From (\ref{D-d2}) and (\ref{d/2ren}) we finally obtain
\be 
\m {\pa \over \pa \m} \car {1 \over (x^2)^{{\frac{1}{2}d} +k}} = 
c_k S_{d-1} (\pa^2)^k \d^{(d)}(x)\,,
\ee
where we used
\be \label{dfunct}
\partial^{2}\frac{1}{(x^2)^{{\frac{1}{2}d}-1}}=-(d-2)S_{d-1}\d^{(d)}(x).
\ee
Therefore, by virtue of (\ref{eq8}), we finally obtain the 
conformal anomaly (\ref{emtr}) as,
\be
\cp_{(2)}^{IJ}=\d^{IJ} C_2(g,\D{=}{\frac{1}{2}d}{+}k) 
{ \p^{\frac{1}{2}d} \over 2^{2k-1} \G(k+1)
  \G(k+\frac{1}{2}d)}\,.\label{conanom} 
\ee
Notice that the ratio $\cp_{(2)}/C_2$ only depends on the conformal
dimension $\D$ and the dimension of spacetime $d$. Since, by assumption,
the operators have vanishing anomalous dimensions, the ratio
does not renormalize.

It is instructive to re-derive this result by using
Ward identities. We will see that one can maintain the 
gravitational Ward identity only at the expense of  
introducing an anomalous term in the trace Ward identity.
The latter is precisely the conformal anomaly.
This analysis also nicely illustrates the need
for specific contact term contributions to correlation
functions.

Symmetries in quantum 
field theory manifest themselves in Ward identities that
Green's functions satisfy.  Let us consider a symmetry 
generated by the  currents $J_\m^\k$, where $\kappa$ denotes the
adjoint representation indices. We take 
the composite operators $\co^I$ to transform as 
\be
\d \co^I = \e^{\kappa}(T^\k)^{IJ} \co^J\,,
\ee 
where $(T^{\k})^{IJ}$ are group generators and $\e^{\kappa}$ arbitrary
parameters. Let $J^I$ denote the
source for the composite operator.  
Then one can, in a standard way \cite{Zinn-Justin}, derive the Ward identity,
\be \label{WI}
\<\pa^\m J^\k_\m(x)\>_s= (T^\k)^{IJ}J^J(x) \<\co^I(x)\>_s\,,
\ee
where the subscript $s$ indicates that the sources have 
not been set to zero. Differentiating (\ref{WI}) a number 
of times with respect to the sources and setting the sources 
to zero, one derives relations among Green's functions
of composite operators. 

Let us consider the case the current $J_\m^\k$ is the dilatation 
current $j_{\m}^{D}(x)=x^{\n}T_{\m\n}(x)$. 
Applying the Ward identity (\ref{WI}) for the case of 2-point functions
we get
\be \label{dilwid}
\<\pa^\m j_\m^D(x_1) \co^I(x_2) \co^J (x_3)\> = 
(d-\D_I) 
(\d^{(d)}(x_{12}) + \d^{(d)}
(x_{13})) {\d^{IJ} C_2(g,\D_I) \over (x_{23}^2)^{\D_I}}
\ee
Since the distribution $1/(x^{2})^{\Delta_I}$ is not well-defined for
$\Delta_{I}=d/2+k$, $k=0,1,2..$,  there may be additional ultralocal
terms on the r.h.s. of the renormalized Ward identity
(\ref{dilwid}). As $j_{\m}^{D}(x)$ is related to 
$T_{\m\n}(x)$, these terms should follow from a proper renormalization
of the 3-pt function $\langle T_{\m\n}\co^{I}\co^{J}\rangle$.  
The conformally invariant 3-point function of
the energy momentum 
tensor and two scalars  is uniquely determined from
conservation, symmetry and tracelessness 
to be
\be
\langle T_{\m\n}(x_{1})\co^{I}(x_{2})\co^{J}(x_{2}) \rangle =
g_{T\co \co}
\frac{\d^{IJ}C_{2}(g,\Delta_I)
}{(x_{12}^{2}x_{13}^{2})^{\frac{1}{2}d-1}(x_{23}^{2})^{\Delta_I
-\frac{1}{2}d+1}}  
\left[X_{\m}(23)X_{\n}(23)-\frac{1}{d}\d_{\m\n}X^{2}(23)\right].\label{TOO}
\ee
with
\be
X_{\m}(23)=\frac{(x_{12})_{\m}}{x_{12}^2}-\frac{(x_{13})_{\m}}{x_{13}^{2}}
\ee
The coupling
constant $g_{T\co \co}$ is fixed from the normalization of the
2-point function 
(\ref{2pt}) alone \cite{Cardy} to be
\be
g_{T\co \co}=\frac{d\Delta_I}{(d-1)S_{d-1}}\,.
\ee
The 3-point functions (\ref{TOO})
satisfies the following Ward identities \cite{OP} corresponding to the fact
that  $T_{\m\n}(x)$  generates Poincar\'e transformations,
dilatations and special conformal transformations  
\bea
\partial^{\m}\langle T_{\m\n}(x_{1})\co^{I}(x_{2})\co^{J}(x_{2})
\rangle  & = &
\partial_{\n}\d^{(d)}(x_{12})\langle\co^{I}(x_{1})\co^{J}(x_{3})\rangle
+\partial_{\n}\d^{(d)}(x_{13}) \langle\co^{J}(x_{1})\co^{I}(x_{2})\,,
\label{emwids} \rangle\\
\langle T^{\m}{}_{\m}(x_{1})\co^{I}(x_{2})\co^{J}(x_{2}) \rangle & = &
(d-\Delta_{I})\left[ \d^{(d)}(x_{12})\langle\co^{I}(x_{1})\co^{J}(x_{3})\rangle
+\d^{(d)}(x_{13}) \langle\co^{J}(x_{1})\co^{I}(x_{2})\rangle\right]\,.\nonumber
\eea
The relations in (\ref{emwids}) should be satisfied by
the fully renormalized 3-pt function (\ref{TOO}).  
Now, (\ref{TOO}) is not well-defined
when $\Delta_I=d/2+k$, $k=0,1,2,..$ as it has short-distance
singularities when two and three points coincide. To remedy this, one
can use differential regularization techniques \cite{diffreg} and write
down an expression for the 3-pt function $\langle
T_{\m\n}\co^{I}\co^{J}\rangle$ which coincides with (\ref{TOO}) for
separate points but is well-defined at the coincident limits
as \cite{OP}
\bea
\langle T_{\m\n}(x_{1})\co^{I}(x_2)\co^J(x_3)\rangle & = &
\frac{d\Delta_{I} \d^{IJ}C_{2}(g,\Delta_{I})}{(d-2)^2
(d-1)S_{d-1}}  
\left(\partial_{\m}\partial_{\n} -\frac{1}{d}\d_{\m\n}\partial^{2}
\right)\left(\frac{1}{(x^2_{12}
    x_{13}^2)^{\frac{1}{2}d-1}(x^2_{23})^{\Delta_{I}  
-\frac{1}{2}d +1}}\right) \nonumber \\ 
& & -\frac{2\Delta_{I}
  \d^{IJ}C_{2}(g,\Delta_{I})}{(d-2)^2 
S_{d-1}}\frac{1}{(x_{23}^{2})^{\Delta_{I}-\frac{1}{2}d+1}} \times \nonu
&&\times
\left[\frac{1}{(x_{13}^2)^{\frac{1}{2}d-1}}\left(\partial_{\m}\partial_{\n} 
  -\frac{1}{d}\d_{\m\n}\partial^{2}\right)
\frac{1}{(x_{12}^2)^{\frac{1}{2}d-1}}
+ (x_{13}^2 \leftrightarrow x_{12}^2) \right] \nonu
& & +\frac{d-\Delta_{I}}{d}\d_{\m\n}[\d^{(d)}(x_{12})+\d^{(d)}(x_{13})]
\d^{IJ}C_{2}(g,\Delta_{I})\car\frac{1}{x_{23}^{2\Delta_{I}}}
\nonumber \\ 
& & +\frac{1}{d}\cp_{(2)}^{IJ}
\d_{\m\n}(\partial^2)^{k}\d^{(d)}(x_{12})\d^{(d)}(x_{13}) 
\label{renTOO} 
\eea
This expression is well-defined for all values of $\Delta_{I}$ when
$\car(1/x^{2\Delta_I})$ is defined by (\ref{D-d2}), (\ref{subtract}). The
last ultralocal term represents the overall ambiguity in regularizing
(\ref{renTOO}). Its coefficient has been chosen
such the gravitational Ward identity is satisfied as we now show.
Taking the divergence of (\ref{renTOO}) we obtain
\bea\label{gravwid}
\partial^{\m}\langle T_{\m\n}(x_{1}) \co^{I}(x_{2})
\co^{J}(x_{3}) \rangle & = & \d^{IJ}C_{2}(d,\Delta_I)
\left[ \partial_{\n}\d^{(d)}(x_{12})\car\frac{1}{x_{13}^{2\Delta_{I}}}
  +\partial_{\n} \d^{(d)}(x_{13})\car\frac{1}{x_{12}^{2\Delta_{I}}}\right]
  \nonumber \\
&&\hspace{-1.7cm} +\frac{1}{d}[\cp_{(2)}^{IJ}-
\m^{2\lambda-d}c_{k}S_{(d-1)}\d^{IJ}C_{2}(g,\Delta_{I})] 
\partial_{\n}[(\partial^2)^k\d^{(d)}(x_{12})\d^{(d)}(x_{13})] \,,
\eea
where we have used (\ref{subtract}). Taking the limit
$\lambda\rightarrow d/2$ in (\ref{gravwid}) we find that the 
last ultralocal term vanishes and the renormalized
gravitational Ward identity is satisfied without an anomalous 
term. The trace Ward identity, however, is now anomalous,
\bea
\langle T_{\m\m}(x_{1})\co^{I}(x_{2})\co^{J}(x_{2}) \rangle & = &
(d-\Delta_I)\left[ \d^{(d)}(x_{12})\langle\co^{I}(x_{1})\co^{J}(x_{3})
  \rangle
+\d^{(d)}(x_{13}) \langle\co^{J}(x_{1})\co^{I}(x_{2})\rangle\right] 
\nonumber\\
& &  + \cp_{(2)}^{IJ}(\partial^2)^{k}\d^{(d)}(x_{12})\d^{(d)}(x_{13})\,,  
\label{danom} 
\eea
which is the desired result.

\subsection{AdS computation}

We now present a calculation of the conformal anomaly using the
adS/CFT correspondence. Our starting point is the following adS$_{d+1}$ action
for massive scalar fields $\phi^I(x_{0},x)$ \footnote{We consider the Euclidean
version  of 
AdS space where  $\rmd \bar{x}^{\m}\rmd \bar{x}_{\m}=\frac{1}{x_{0}^{2}}
(\rmd x_{0}\rmd 
x_{0}+\rmd x^{i}\rmd x^{i})$, with $i=1,2,3,4.$ and
$\bar{x}^{\m}=(x_{0},x_{i})$. 
The boundary of this space is the sphere
{\bf{S}}$^{\rm d}$ consisting of
{\bf{R}}$^{\rm d}$ at $x_{0}=0$ and a single point at $x_{0}=\infty$.}
with cubic vertices  
\beq
I[\phi^I,g] = \int\rmd^{d+1}x\sqrt{g}\left[ \frac{1}{2}\left
    ( \partial_{\m}\phi^I\partial^{\m}\phi^I+m_I^2\phi^I\phi^I\right)
  +\frac{1}{3!}w^{IJK}\phi^{I}\phi^{J}\phi^{K}\right]\,,\label{Lagr} 
\eeq
where $w^{IJK}$ are some coupling constants. Following
\cite{Freedman2,Viswa} we impose the Dirichlet boundary conditions 
\beq
\lim_{x_{0}\rightarrow
\e}\phi^{I}(x_{0},x)=\e^{d-\Delta_{I}}\phi_{0}^{I}(x)\,,
\label{bcond}
\eeq
at the adS boundary $x_{0}=\e\ll 1$. This yields the $\e$-dependent
generating functional for the boundary 2-point and 3-point functions 
as 
\bea
W_{\e}[\phi_{0}^I] & = &
\frac{1}{2}\int\rmd^{d}x_1\,\rmd^{d}x_2\,\phi^{I}_{0}(x_1)\phi^{J}_{0}(x_2)
G_{(2)}^{IJ}(x,\Delta_{I};\e) \nonumber \\
& & +\frac{1}{3!}\int\rmd^{d}x_{1}\rmd^{d}x_{2} \rmd^{d}x_{3}
\phi^{I}_{0} (x_{1}) \phi^{J}_{0}(x_{2})
\phi^{K}_{0}(x_{3})G_{(3)}^{IJK}(x_i,\Delta_i;\e)\,,\label{genfunct} \\ 
G_{(2)}^{IJ}(x,\Delta_{I};\e) & = & \d^{IJ}
\int\frac{\rmd^{d}p}{(2\pi)^{d}}e^{ipx_{12}} 
\frac{1}{\e^{2k_I}}\left[\left(\frac{1}{2}d-k_I\right)-|p|\e\frac{
\ck_{k_I -1}(|p|\e)}{\ck_{k_I}(|p|\e)}\right]\,,\label{ads2pt}\\
G_{(3)}^{IJK}(x_i,\Delta_i;\e) & = & w^{IJK}\int\frac{\rmd^{d}p}{(2\pi)^{d}}
\frac{\rmd^{d}q}{(2\pi)^{d}} e^{-{\rm 
i}px_{13}-{\rm i}qx_{23}}
{\cal{I}}_{3}(p,q,\Delta;\e) \,,\label{ads3pt}\\ 
{\cal{I}}_{3}(p,q,\Delta;\e) & = & \int_{\e}^{\infty}\rmd
x_{0}\,x_{0}^{\frac{d}{2}-1}
\frac{\ck_{k_{I}}(|p|x_{0}) \ck_{k_{J}}(|q|x_{0})
\ck_{k_{K}}(|p+q|x_{0})}{\ck_{k_{I}}(|p|\e)\ck_{k_{J}}(|q|\e)
\ck_{k_{K}}(|p+q|\e)}\,
\epsilon^{-(k_{I}+k_{J}+k_{K})}\,,\label{ads3int}
\eea 
where the massive scalar bulk fields $\phi^{I}(x_{0},x)$ 
have adS-masses and dimensions respectively 
\beq
m_{I,J,K}^{2}=k_{I,J,K}^{2}-\frac{1}{4}d^{2}\,,\,\,\,\,\,
\Delta_{I,J,K}=\frac{1}{2}d +k_{I,J,K} \label{adsmass}
\eeq
For $k_{I,J,K}$ integers one deals with Kaluza-Klein states.
The dimensions of the operators that form the 2- and 3-point functions
in (\ref{genfunct}) satisfy the condition
(\ref{eq9}) and they contribute to the conformal anomaly of the
boundary CFT. 

The inverse of the infinitesimal distance $1/\e$ from the adS boundary
can be used as 
an UV mass-scale $\m$ in the boundary theory \cite{SuWi} in accordance
with the ideas of holography \cite{Hooft}. In the limit $\m\rightarrow \infty$
($\e\rightarrow 0$) one then obtains from (\ref{ads2pt}) and (\ref{ads3pt}) the
renormalized 2- and 3-point functions of the boundary
CFT. However, the limit $\e\rightarrow 0$ is a delicate procedure as
it is related to coincident singularities which are in turn related to
conformal anomalies. In particular, if there exist the logarithmic terms in
(\ref{genfunct}) in the limit $\e\rightarrow 0$ then, from
(\ref{eq8}), these would give rise to conformal anomalies.

It is relatively simple to obtain the $\m$-dependence for 2-point
functions. To this end we observe that for integer $k_I$ the $\ck$-Bessel
functions in (\ref{ads2pt}) have logarithmic terms as
$\e\rightarrow 0$ \cite{Gradshteyn}. These terms give rise to
$\d$-function terms upon 
evaluating the $\m$-derivative of the generating functional,
which in turn yield a non-vanishing trace
anomaly  in (\ref{emtr}) as  
\beq
{\cal{P}}_{(2)}^{IJ}=\d^{IJ}C_2
\frac{\pi^{\frac{1}{2}d}}{2^{2k_I -
1} \Gamma(k_I +1)\Gamma(\frac{1}{2}d +k_I)}\,,\,\,\,\,\,\,\,\, C_{2} =
\frac{2k_{I}\Gamma(\frac{1}{2}d+k_I)}
{ \pi^{\frac{1}{2}d}\Gamma(k_{I})} \label{tranom} 
\eeq
This coincides with (\ref{conanom}) since $C_{2}$ is the 2-point
function normalization obtained in \cite{Freedman2,Viswa}.

\section{Non-Renormalization Theorems in $\cn=4$ SYM}

We comment in this section on the application of our results to 
$\cn=4$ SYM theory.
$\cn=4$ SYM theory is most succinctly described in harmonic 
superspace \cite{HH,HW1}. The YM multiplet, i.e. 6 real 
scalars $\f^i$ in the ${\bf 6}$ of the global symmetry group $SU(4)$,
4 Weyl fermions $\l_\a^A$ in the ${\bf 4}$ of $SU(4)$  and a vector field
$A_\m$, is described by
an analytic superfield $W$ taking values in the adjoint of 
$SU(N)$.
$W$ is a covariantly analytic superfield. The explicit form
of the constraints can be found, for example, in \cite{HW1,Fer2}.
As it was shown in \cite{Fer2} (see also \cite{Fer1,Fer3})
the short multiplets of $\cn=4$ 
SYM have a description in terms of the analytic superfields $A_p = \Tr W^p$,
where the trace is over the color indices. The lowest 
component of this superfield is 
$A^{i_1...i_p}=\Tr \f^{\{i_1}...\f^{i_p\}}$-traces.
In \cite{HSW,EHW}, 2- and 3-point function were determined 
directly in analytic superspace by solving the 
$\cn=4$ superconformal Ward identities. It has been 
established there that these Ward identities uniquely determine
all 2-point function and 3-point functions up to a constant.
Therefore, it is sufficient to know the correlation functions 
of $A^{i_1...i_p}$ in order to determine the correlation 
functions of all $\cn=4$ chiral primary operators.
For these operators, our theorem is directly applicable.

In a number of recent papers \cite{Intr1,Intr2,EHW}, the
non-renormalizability of 
2- and 3-point functions of chiral primary operators
was linked to a bonus $U(1)_Y$ symmetry 
that $\cn=4$ SYM is expected to have 
in a specific limit. However, the arguments
in \cite{Intr1,Intr2} did not take into account 
possible contact terms in correlation functions. Such local terms 
do not respect the bonus $U(1)_Y$, 
so even if the selection rules advocated in 
\cite{Intr1,Intr2} hold (away from coincidence points),
this does not imply non-renormalizability 
of 2- and 3-point functions of chiral primary operators.
However, such conclusion would be correct if one 
could show that the local terms saturate by 
tree-level contractions. 

Let us review the argument of \cite{Intr1}. 
The adS/CFT duality connects $\cn=4$ $SU(N)$
SYM with string theory on $adS_5 \times S^5$. The
latter has a non-perturbative $SL(2,Z)$ symmetry. 
This symmetry is enhanced to $SL(2,R)$ in the
limit where supergravity is valid. In particular,
the theory has an extra $U(1)_Y$ symmetry, the 
latter being the maximal compact subgroup of $SL(2,R)$.
String corrections and higher dimension operators break this symmetry,
so the bonus $U(1)_Y$ cannot be, and it is not, 
a symmetry of $\cn=4$ SYM theory. However, operators can be assigned a 
definite $U(1)_Y$ charge:
$F_{\a\b}, \psi$, and $\f$ have $U(1)_Y$ charge equal to -2, -1 and 
0, respectively ($F_{\a \b}$ is the anti-self dual part of the
field strength, see below). The complex conjugate fields have opposite 
$U(1)_Y$ charge. It was argued in \cite{Intr1} and 
perturbatively checked in a number of cases \cite{Intr2},
that 2- and 3-point functions exactly satisfy a
selection rule that follows from $U(1)_Y$.
In particular, all 2-point and 3-point functions 
with at least two protected operators 
can be non-zero only if their charges  
sum up to zero. However, 
contact term contribution to the 
correlation function need not respect the $U(1)_Y$ selection 
rule.

Let us consider the pure Yang-Mills sector 
of the $\cn=4$ theory for the moment. The quadratic part of the action reads
\bea \label{free}
\cl&=&\Tr [{1 \over 2 g_{YM}^2} F^{a b} F_{a b} 
+ {\theta_{YM} \over 16 \p^2} F^{a b} \tilde{F}_{a b}]
\nonu
&=&\Tr \left[{i \over 4 \p}\left( -\t F^{\a \b} F_{\a \b} 
+ \bar{\t} F^{\dot{\a} \dot{\b}} F_{\dot{\a} \dot{\b}}\right) \right] 
\eea
where $F^A_{a b}=\pa_a A^A_b - \pa_b A^A_a$ denotes the quadratic 
part of the field strength,  
$\t \equiv {\theta_{YM} \over 2 \p} + {4 \p i \over g_{YM}^2}$,  
$\tilde{F}_{a b}=\half \e_{a b c d} F^{c d}$ and 
$\Tr T^A T^B = - \half \d^{AB}$. We use the conventions of 
\cite{superspace}. 
In particular, the field strength decomposes into 
self-dual and anti-self dual pieces as
$F_{a b}=C_{\a \b} F_{\dot{\a} \dot{\b}} +C_{\dot{\a} \dot{\b}} F_{\a \b}$.
$F_{\dot{\a} \dot{\b}}$ and $F_{\a \b}$ are symmetric in their 
indices. $C_{\a \b}$, the antisymmetric invariant tensor 
of $sl_2$, is used to raise and lower indices
using the north-east convention for both undotted and 
dotted indices. 
One can easily derive the following Ward identity
\be \label{tauWI}
{\pa \over \pa \t} \<\prod_{i=1}^n \co_i(x_i)\> =
{1 \over 8\p} \int d^4z \<\co_\t(z) \prod_{i=1}^n \co_i(x_i)\>
\ee
where $\co_\t = (8\p i) \pa \cl/ \pa \t = F^{\a \b A} F_{\a \b}{}^A + ...$. 
The dots indicate contributions from the other fields
and the interaction terms. We will be concerned with the free field
contributions, so the interaction terms are not relevant. 
Furthermore, with appropriate normalization of the fermion 
and scalar fields their kinetic term is independent of 
$\t$. With such normalization, there are no extra terms
in $\co_\t$. 

The argument of \cite{Intr1} has as follows.
According to the selection rules proposed there,
if the correlation function in the l.h.s.
is non-zero, then the correlation function in the 
r.h.s. is necessarily zero since, 
because the operator $\co_\t$ has $U(1)_Y$ charge
equal to -4, it cannot be that in both sides the 
$U(1)_Y$ charges sum up to zero.
However, the selection rules can be, and actually are violated
by contact terms (as it has already been observed in \cite{Intr2}).
One can ignore contact terms in the l.h.s. by taking all operators
at separate points, but this cannot be done in the r.h.s as 
one integrates over the insertion point of $\co_\t(z)$. 
In fact, the tree-level contributions to these contact terms 
complete the $\t$-derivative into a modular covariant derivative.

We now calculate these local terms.
In the Feynman gauge, the gauge field propagator reads
\be
\<A^A_{\a \dda}(x_1) A^B_{\b \ddb}(x_2)\> = i g_{YM}^2 \d^{AB}
C_{\a \b} C_{\dda \ddb} \D(x_1-x_2)
\ee
where $\D(x_1-x_2)$ satisfies $\Box \D(x_1-x_2)=\d(x_1-x_2)$.
From here it follows that 
\bea \label{comr}
&&\<F^A_{\a \b}(x_1) F^B_{\g \d}(x_2)\>=i {g_{YM}^2 \over 4} \d^{AB}
(C_{\b \g} C_{\a \d} + C_{\a \g} C_{\b \d}) \d^{(4)}(x_1-x_2) \nonu
&&\<F^A_{\a \b}(x_1) F^B_{\ddg \ddd}(x_2)\>=i {g_{YM}^2 \over 2}\d^{AB}
(\pa_{\a \ddd} \pa_{\b \ddg} + \pa_{\a \ddg} \pa_{\b \ddd} )
\D(x_1-x_2)
\eea

From (\ref{comr}) follows that tree-level contractions  
between the operator $\co_\t$ and some other operators $\co$
simply counts, up to the factor of $i g_{YM}^2$, the number of 
$F_{\a \b}$ that the operator contains. In particular,
\be \label{trr}
\int d^4 z \< \co_{\t}(z) \prod_{i=1}^p \co_{\t}(x_i)
\prod_{i=1}^q \co_{\bar{\t}}(y_i) \> =
i g_{YM}^2 (2 p) \<\prod_{i=1}^p \co_{\t}(x_i)
\prod_{j=1}^q \co_{\bar{\t}}(y_j) \>
\ee
If we take all $x_i$ and $y_i$ to be at separate points, the 
tree-level contribution to the correlation function is non-
zero only for $p=q$. In our discussion below
we always take the external legs to be at separate 
points. The tree-level contribution to (\ref{tauWI}) reads
\be \label{mod}
\left({\pa \over \pa \t}  - i {2p \over 2 \t_2} \right)
\<\prod_{i=1}^p \co_{\t}(x_i)
\prod_{i=j}^q \co_{\bar{\t}}(y_j) \>|_{{\rm tree}} =0
\ee
Before we proceed let us present an example.
Consider the case of the 2-point function between 
$\co_\t$ and $\co_{\bar{\t}}$. Then (\ref{mod}) gives
the following differential equations for the arbitrary 
coefficient $C_2^{\t \bar{\t}}$ of the 2-point function,
\be
\pa_\t C_2^{\t \bar{\t}} = {i \over \t_2} C_2^{\t \bar{\t}}.
\ee
It is easy to integrate this equation. The solution is that 
$C_2^{\t \bar{\t}} \sim (g^2_{YM})^2$, which is 
what one expects for a tree-level contribution.

Going back to (\ref{mod}),
we observe that the combination $\cd_\t=i(\pa_\t -i{2p \over 2 \t_2})$
is the modular covariant derivative that maps a $(2p,q)$ 
modular form\footnote{
A non-holomorphic modular form $F^{(p, q)}$ of weight $(p,q)$ transforms as,
$F^{(p, q)} \to F^{(p, q)} (c \t +d)^p (c \bar{\t} +d)^q$, under 
the $SL(2,Z)$ transformation, $\t \to {a \t + b \over c \t + d}, 
ad-bc=1, a,b,c,d$ integers.} to a $(2p+2,q)$ form. 
A similar calculation that involves $\co_{\bar{\t}}$ yields
\be \label{mod1}
\left({\pa \over \pa \bar{\t}}  + i {2q \over 2 \t_2} \right)
\<\prod_{i=1}^p \co_{\t}(x_i)
\prod_{j=1}^q \co_{\bar{\t}}(y_j) \>|_{{\rm tree}} =0
\ee
The modular covariant derivatives act in a way which is 
consistent with the assignment of weight $(2,0)$ to 
every factor of $\co_\t$ present in the correlation 
function, and weight $(0,2)$ for every factor of $\co_{\bar{\t}}$.
The relation (\ref{mod}), (\ref{mod1}) can be compactly
expressed as (with the understanding that 
only correlation function with external legs
at separate points are considered)
\be \label{Zcov}
\cd_\t Z[J,\t] = \cd_{\bar{\t}} Z[J,\t] =0,
\ee
where $J$ collectively denotes the 
sources for the composite operators.
It may seem surprising at first that all correlation functions 
are covariantly constant w.r.t. the modular group.
However, this has a rather simple explanation. 
The calculation only uses data that come from 
the quadratic action (\ref{free}). This action 
is identical to the action describing $N{-}1$ copies of Maxwell fields.  
The relations (\ref{Zcov}) may be easily shown to hold by noting that 
(i) the $\theta_{YM}$ is trivial since we consider the theory
on $R^4$, (ii) the $g_{YM}$ dependence is also trivial since 
one can always rescale away the coupling constant from a 
free-theory. In the presence of sources, the source terms 
will acquire some $g_{YM}^2$ dependence after 
the coupling constant has been rescaled away from the kinetic term. The 
connection piece in (\ref{Zcov}) simply cancels these terms.
It follows that by considering appropriately normalized 
operators (in our case $\co_\t \to \co_\t' = \co_{\t} \t_2$) 
one can completely scale away the coupling constant from 
both the kinetic and the source term. Then  
the connection piece in (\ref{Zcov}) vanishes as well. Since $\t_2$ transforms
as $(-1,-1)$ form, $\co_\t'$ transforms as $(1,-1)$ 
form which are also the weights assigned in \cite{Intr1}.
Incorporating the scalars and fermions in the discussion
does not bring any qualitative change.

To summarize our discussion: for a free theory,
one can rescale away the coupling constant, and then 
all correlation functions of (appropriately 
normalized) operators are independent of the coupling constant.
In an interacting theory,
although one can scale away the coupling constant
from the propagators, the coupling constant will 
appear through the interactions. This means 
that one can always arrange, by considering 
appropriately normalized operators,
such that the tree-level contact terms in (\ref{trr})
vanish. The non-trivial content of 
(\ref{trr}) is in possible loop-contributions.
In particular, non-renormalizability of 2- and 3-point 
functions (or, more generally, of n-point functions) 
is equivalent to saturation of (\ref{trr})
by free fields. This is not guaranteed by 
the bonus $U(1)_Y$ symmetry because the
selection rules are not respected
by contact terms.

\section{Summary and Outlook}

We have presented a non-renormalization theorem for the conformal anomaly
in the presence of external sources for operators with vanishing
anomalous dimensions. As an illustrative example 
we computed the conformal anomaly associated with 2-point functions,
both directly from CFT and also from the adS/CFT correspondence and 
found agreement. The
corresponding 3-point function calculations will be reported elsewhere
\cite{PS}. Our results imply that (i) if the usual relation between the
divergence of the $R$-current and the trace anomaly in supersymmetric theories
is valid in the presence of external sources and (ii) if the former
saturates by free-fields, then  the 3-point functions of operators
with protected dimensions are non-renormalized. These two conditions,
although seemingly a direct generalization of the ingredients 
entering the known
supersymmetric non-renormalization theorem of \cite{Freed}, they have not been
discussed in the literature, and therefore they are not guaranteed. 

We have also discussed the recent proposal of \cite{Intr1} relating the
non-renormalizability of 2- and 3-point functions to a bonus $U(1)_Y$
symmetry of $\cn=4$ SYM. Such a proposal does not take into
account possible ultralocal terms that may appear in conformally
invariant $n$-point functions in coincident limits. Nevertheless, the
conclusions drawn in \cite{Intr1,Intr2,EHW} would be correct if these
contact terms saturate by free-fields.

Regarding further development of our work, it is essential that the
``generalized Adler-Bardeen'' theorem is studied in the presence of
external fields in both ${\cal N}=4$ SYM and also on ${\cal N}=1,2$
theories. Also important would be to understand the contact terms in
$U(1)_Y$ invariant 
2-, 3-point functions and to see whether they saturate by free fields.
To this respect, we note that the finiteness of the $\cn=4$ SYM 
may play a crucial role in such analysis. For instance, the operator $\co_\t$,
being a derivative of the Lagrangian, does not renormalize. 
Notice also that the operator $\co_\t$ is in the same supersymmetry multiplet 
with the energy-momentum tensor and the R-currents, so 
the answers to the above two questions may be related.

\section*{Acknowledgements}
We would like to thank Robbert Dijkgraaf and Erik Verlinde
for useful discussions. K.S. would like to thank Nordita and 
the Niels Bohr Institute for hospitality and support 
during the final stage of this project.
K.S. is supported by the Netherlands Organization for Scientific 
Research (NWO). A.P. is supported by the Greek National Scholarships
Foundation (I.K.Y.).

\end{document}